\documentclass[twocolumn,showpacs,preprintnumbers,amsmath,amssymb]{revtex4}

\usepackage{epsfig}
\usepackage{subfigure}
\usepackage{graphicx}
\usepackage{dcolumn}
\usepackage{bm}

\preprint{APS/123-QED}

\begin{document}
\title{Spin dependent electron transport through a magnetic resonant
tunneling diode}
\author{P. Havu, N. Tuomisto, R. V\"a\"an\"anen, M. J. Puska, and
R. M. Nieminen} 
\affiliation{
Laboratory of Physics, Helsinki University of Technology,
P.O. Box 1100, FIN-02015 HUT, Finland }

\date{\today}

\begin{abstract}
Electron transport properties in nanostructures can be modeled, for
example, by using the semiclassical Wigner formalism or the quantum
mechanical Green's functions formalism. We compare the performance and
the results of these methods in the case of magnetic
resonant-tunneling diodes. We have implemented the two methods within
the self-consistent spin-density-functional theory. Our numerical
implementation of the Wigner formalism is based on the
finite-difference scheme whereas for the Green's function formalism
the finite-element method is used. As a specific application, we
consider the device studied by Slobodskyy {\it et al.}
[Phys. Rev. Lett. {\bf 90}, 246601 (2003)] and analyze their
experimental results. The Wigner and Green's functions formalisms give
similar electron densities and potentials but, surprisingly, the
former method requires much more computer resources in order to obtain
numerically accurate results for currents. Both of the formalisms can
successfully be used to model magnetic resonant tunneling diode
structures.
\end{abstract}


\maketitle

\section{Introduction}

The future spintronics technology requires controlled spin injection
into semiconductor materials. The problem can be solved using
different techniques and constructions (see, for example,
Refs. \cite{inj1,inj2,inj3,inj4,inj5}).  The magnetic resonant
tunneling diode (RTD) demonstrated by Slobodskyy {\it et al.}
\cite{mit} is one of the most promising solutions. Their magnetic RTD
is based on the quantum well made of dilute magnetic semiconductor
ZnMnSe between two ZnBeSe barriers and surrounded by highly $n$-type
ZnSe. In the presence of a magnetic field there is a giant Zeeman
splitting between the spin-up and spin-down electron states in the
quantum well region. The corresponding resonance peaks in the electron
current separate as a function of the bias voltage, and the device
controls the electron spin states using the bias voltage rather than
external magnetic fields. This is a useful property for possible
practical applications.

In a typical nano-scale transport problem two or more electrodes are
connected to a functional nanostructure. An important nano-system is
the RTD composed of layers of semiconducting materials.  The offset
between the band edges of the materials give rise to the two potential
barriers seen by carriers. In the quantum well between the barriers,
resonance states with finite energy width and enhanced amplitude are
formed. The electron current increases rapidly when a resonance state
appears in the conducting window, determined by the overlap of the
occupied source and unoccupied drain conduction electron bands. When
the resonance state drops with increasing bias voltage below the
source conduction band, the current diminishes causing a region of
negative resistance.  Besides being technologically interesting the
RTD is also important as a simple test case for different formalisms
and computational schemes.

Theoretical modeling and computational simulation are essential for
the development of functional nanostructures.  Electron transport
properties can be modeled using different formalisms at different
levels of sophistication.  Two methods widely used are the Wigner
function (WF) \cite{frensley,jacoboni} and the Green's functions (GF)
formalism \cite{datta}.  The WF approach is a semiclassical transport
formalism that enables the study of systems exhibiting quantum
interference and tunneling effects. The electron density and the
current are obtained from the Wigner function, which is in turn
calculated by solving the Liouville--von~Neumann equation. The GF
formalism is a fully quantum mechanical scheme with an increased
complexity with respect to the WF formalism. Both the WF and GF
formalisms enable self-consistent non-equilibrium calculations
corresponding to a finite bias voltage between the electrodes.

RTD's can be modeled as structures which are translationally invariant
parallel to the layers. This makes the computations one-dimensional.
The WF and GF formalisms are very popular schemes in their modeling
(see, for example, \cite{g1,g2,g3,w1}).  In this work we study the
feasibility of the WF and GF formalisms within the context of the
self-consistent spin-density-functional theory (SDFT) and its local
spin-density approximation to model magnetic RTD's. Our implementation
of the WF formalism is based on the usual discretation of the partial
differential equations on position and momentum point grids. Our
implementation of the GF formalism employs the finite-element method
(FEM). We have already published our FEM scheme for two-dimensional
nanostructures \cite{oma} and used it in applications \cite{oma2}.  In
this work we critically compare the performance and the results of the
WF and GF implementations, for quasi-one-dimensional RTDs. Moreover, we
choose the structure parameters of our test system to correspond the
magnetic RTD device by Slobodskyy {\it et al.}  \cite{mit}. Thus, we
can compare our results also with experiments and actually analyze the
results of the measurements.

Below we use the effective atomic units which are derived by setting
the fundamental constants $e = \hbar = m_e = 1$, and the material
constants $m^* = \epsilon = 1$. $m^*$ and $\epsilon$ are,
respectively, the relative effective electron mass and the relative
dielectric constant to be used in the effective mass approximation. For
ZnSe $m^* = 0.145$ and $\epsilon = 9.1$ \cite{me,me2}. We have used
same values also for the other materials of the magnetic RTD.  The
effective atomic units can be transformed to the usual atomic and SI
units using the relations

\begin{center}
\begin{tabular}{l l l l}
Length: & $1 \, a_0^*$ & $= 1 \frac{\epsilon}{m^*} a_0$ &$\approx$ 3.32 nm\\ 
Energy: & $1 \, {\rm Ha^*}$ &$= 1
\frac{m^*}{\epsilon^2} {\rm Ha}$ & $\approx$ 47.6 meV \\ 
Current: &
$1 \, {\rm a.u.}^*$&$ = 1 \frac{m^*}{\epsilon^2} {\rm a.u.} $ &
$\approx$ 11.6 $\mu$A.
\end{tabular}
\end{center}
Above, Ha denotes the Hartree energy unit.

In Sec. II we introduce the model for the magnetic RTD and explain the
use of the SDFT in the calculations. In Sec. III we briefly explain
the GF and WF formalisms. In Sec. IV we give the results of the
comparisons between the two formalisms and compare the calculated and
measured results and obtain information on the electronic structure of
the device in question. Sec. V contains our conclusions.

\section{Model}

\subsection{Structure of magnetic RTD}

\begin{figure}[!htb]
\begin{center}
\epsfig{file=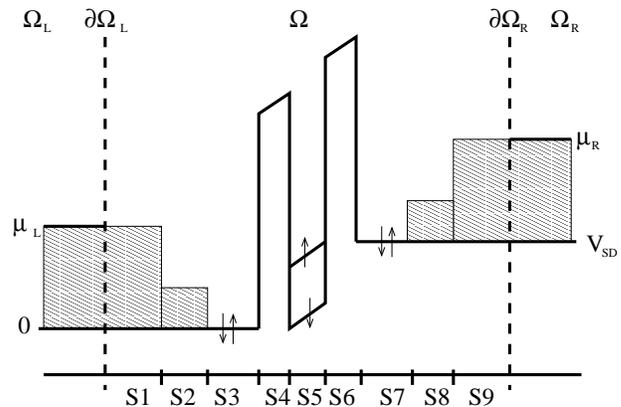,width=0.45\textwidth}
\end{center}
\caption{\label{tunneli} Magnetic RTD model. The shaded areas denote
positive background charge. The detailed information about the
different layers S1,...,S9 are given in Table~\ref{Sx}.}
\end{figure}

The model for the magnetic RTD is shown in Fig.~\ref{tunneli}. We
assume the semiconducting layers to be infinitely wide in the lateral
directions so that the system is translationally invariant in the
direction perpendicular to the current. The doped regions (shadowed
areas in Fig.~\ref{tunneli}) are modeled by a uniform positive
background charge. The potential barriers due to the
discontinuity of the conduction band between two materials are
described by constant external potentials.

The system is divided into the central region $\Omega$ and the outside
regions $\Omega_L$ and $\Omega_R$. All the structural variations and
interesting phenomena take place in region $\Omega$ which is chosen
large enough so that the effect of the RTD device on the electron
density has vanished at the boundaries $\partial \Omega_{L/R}$.
$\Omega_{L/R}$ are the semi-infinite leads where the electron density
and the potential are constant. The bottom of the conduction band and
the Fermi-level $\mu_{R}$ in the right lead are shifted by the bias
voltage $V_{SD}$ with respect to the corresponding values in the left
lead (see Fig.~\ref{tunneli}). In our model the electron transport is
ballistic with no phonon or defect scattering. This means that the
total potential drop takes place within $\Omega$.

\begin{table}[b]
\begin{center}
 \caption{Parameters of the different layers S1,...,S9 used in the
   calculations.}
\begin{ruledtabular}
\begin{tabular}{|c|c|c|c|c|} 
Region & Material & Width & Doping level & Potential $V_{W}$  \\
& & (nm) & (cm$^{-3}$) & (meV)\\
\hline
S1, S9 & ZnSe                     & 25 & 15$\times 10^{18}$ &  0      \\ 
S2, S8 & Zn$_{0.97}$Be$_{0.03}$Se & 15 & 1$\times 10^{18}$  &  0      \\ 
S3, S7 & ZnSe(i)                  & 10 &  0                 &  0      \\ 
S4, S6 & Zn$_{0.7}$Be$_{0.3}$Se   & 5  &  0                 & 92      \\
S5     & Zn$_{0.96}$Mn$_{0.08}$Se & 9  &  0  & $\pm \frac{1}{2}\Delta E$  \\ 
\end{tabular}\label{Sx}
\end{ruledtabular}
\end{center}
\end{table}

In addition, $\Omega$ is divided into nine smaller parts S1, ... S9 as
shown in Fig.~\ref{tunneli}. These regions describe different
semiconductor material layers. The parameters of the layers are given
in Table~\ref{Sx}. We have chosen the widths and the doping densities
of our magnetic RTD structure similar to those in the actual device
made by Slobodskyy {\it et al.}  \cite{mit}.  The positive background
charge densities in regions S1 and S9 are equal to that in
$\Omega_{L/R}$, in S2 and S8 they are much smaller, and next to the
potential barriers there is no background charge at all.

The regions S4 and S6 are the potential barriers. Because there is no
definitive information about the barrier height we estimate it by
comparing the shapes of the calculated current-voltage curves to the
measured ones. By the shape we mean mainly the widths of the
resonances; the positions of the resonances are quite insensitive to
the barrier height.  The barrier height is the only structural
parameter which we have to determine by fitting. We find that the
barrier height of 23\% of the band gap difference
\cite{aukko-offset,aukko-offset2} between ZnSe and ZnBeSe results in a
good fit.

The quantum well S5 is made of the dilute magnetic semiconductor
ZnMnSe. An external magnetic field causes a giant Zeeman effect,
splitting the spin-up and spin-down electron states.  In the other
parts of the RTD the spin splitting is small and assumed to vanish. We
calculate the spin splitting $\Delta E$ in S5 as Slobodskyy {\it et
al.} \cite{mit}, i.e.
\begin{equation}\label{deltaE}
\Delta E = N_0 \alpha x s_0 B_s \left( \frac{s g \mu_B B}{k_B
  (T+T_{eff})} \right).
\end{equation}
Above, $N_0 \alpha$ is the s-d exchange integral, $x$ the Mn
concentration, $g$ is the Land\'e factor, $\mu_B$ the Bohr magneton,
$B_s$ is the Brillouin function of spin $s$, $s_0$ is the effective Mn
spin and $T_{eff}$ is the effective temperature. The values of the
parameters are $N_0 \alpha$ = 0.26 eV, $x$ = 8\%, $s$ = 5/2, $g$ =
2.00, $s_0$ = 1.13 and $T_{eff}$ = 2.24 K. The values of the $\Delta E$
for the relevant magnetic fields and temperatures (Sec.~V)
are collected into Table~\ref{spitting}.

\begin{table}
\caption{Values of the spin splitting $\Delta E$ in the cases 
considered in Sec.~V.} 
\begin{ruledtabular}
\begin{tabular}{|c|c|c|}
 Magnetic field (T)& Temperature (K)& $\Delta E$ (meV)\\
\hline
0 & 4.2   & 0.0  \\
2 & 4.2   & 10.4  \\
4 & 4.2   & 16.7  \\
6 & 4.2   & 19.8  \\
6 & 0     & 23.2  \\
6 & 8     & 16.2  \\
6 & 30    & 6.7  \\
\end{tabular}\label{spitting}
\end{ruledtabular}
\end{table}

\subsection{Spin-density-functional theory}

In order to model electron-electron interactions we use the SDFT
within the local-density approximation. The electronic structures and
currents are calculated using the GF or the WF formalisms, which we
will explain later in Sec. III. In both formalisms the spin-up
($\sigma= \uparrow$) and spin-down ($\sigma= \downarrow$) electron
densities $n_{\uparrow,\downarrow}(x)$ correspond to the effective
potentials
\begin{equation}\label{veff}
V_{eff}^\sigma(x) = V_C(x) + V_{xc}^\sigma(x) + V_{W}^\sigma(x),
\end{equation}
where $V_C$ is the Coulomb potential, $V_{xc}^\sigma$ the
exchange-correlation potential \cite{xc} and $V_{W}^\sigma$ the
external potential including the barriers and the giant Zeeman spin
splitting $\Delta E$ (see Table~\ref{spitting}). The spin densities
and the effective potentials are solved self-consistently.

The Coulomb potential $V_C$ is calculated using the modified Poisson 
equation \cite{poisson}
\begin{equation}\label{poisson}
\begin{aligned}
\nabla^2 V_C^{i+1}(x) - &k^2 V_C^{i+1}(x) \\
 =& -4 \pi [n_+(x) - n_-^i(x)] - k^2
V^i_{C}(x),
\end{aligned}
\end{equation}
where $n_-(x)= n_{\uparrow}(x)+ n_{\downarrow}(x)$ is the total
electron density and $n_+(x)$ is the positive background charge.
Index $i$ counts the self-consistency iterations, so that $V^i_{C}(x)$
is the solution from the previous self-consistency iteration. Above,
$k$ is a parameter which controls the screening of the potential
fluctuations due to the charge sloshing between the iterations. A
reasonable choice of the $k$ parameter is of the order of the
Thomas-Fermi wave vector, in which case the solution does not depend
on $k$ and the number of the self-consistency iterations needed is
remarkably reduced.  Besides, the stability obtained by the use of
the modified Poisson equation, we stabilize the iterations also by
mixing the old effective potential $V^i_{eff}$ with that obtained from
Eqs. (\ref{veff}) and (\ref{poisson}).  I.e.,
\begin{equation}
V^{i+1}_{eff} = \alpha V_{eff} +  (1-\alpha) V^i_{eff},
\end{equation}
where the feedback parameter $\alpha$ is typically 0.2 in our calculations.

We calculate the currents due to the spin-up and spin-down electrons
through the magnetic RTD as a function of the voltage. The calculation
always starts from the zero bias voltage $V_{SD}$. When the
self-consistent non-biased (equilibrium) result is reached, we
increase $V_{SD}$ in small steps and iterate at every value until
convergence.  The effective potential corresponding to the previous
$V_{SD}$ value is used as the starting point of the iterations. This
ensures the stability of the process. Actually we found that the
resulting current voltage curve does not depend on the sweep direction
of the bias voltage. This is due to the very small electron density in
the quantum well region S5 even in the case of occupied resonance
states. This insensitivity is in accord with the findings by Slobodskyy
{\it et al.}

\section{Formalisms}

In this section we present the GF and WF formalisms by ignoring the
spin-dependence for simplicity. The generalization to the
spin-dependent forms, which we use in the actual calculations, is
straightforward.

\subsection{Green's function formalism}

The GF formalism used in the electron density and transport
calculations is explained in detail in Ref. \cite{datta}.  We have
implemented this formalism using the finite element method (FEM). Our
FEM formulation for two-dimensional nanostructures is discussed in
Ref. \cite{oma}.

Our present magnetic RTD system is translationally invariant in
directions perpendicular to the electron current. The total electron
energy $\omega_{tot}$ can then be divided into two parts
\begin{equation}
\omega_{tot} = \omega + \omega_{\bot},
\end{equation}
where $\omega_\bot$ is the kinetic energy in the perpendicular
directions and $\omega$ includes the kinetic energy along the current
and the (one-dimensional) potential energy. Now we can write a
one-dimensional equation for the single-particle retarded Green's
function in the real space using spatial coordinates, $x$ and $x'$ in
the direction perpendicular to the layers:
\begin{equation}\label{greenR}
\big{(}\omega-\hat{H} \big{)}
G^r(x,x';\omega) = \delta(x-x'),
\end{equation}
where $\hat{H}$ is the Hamiltonian
\begin{equation}\label{hamiltonian}
\hat{H} = -\frac{1}{2}\nabla^2 + V_{eff}(x).
\end{equation}
Above, $\omega$ has a small imaginary part, i.e., $\omega = \omega' +
i \eta$. This distinguishes between the retarded and the advanced
Green's functions. Eq.~(\ref{greenR}) is solved using open boundary
conditions at $\partial \Omega_L$ and $\partial \Omega_R$. This means
that electron wave functions penetrate the boundaries without
reflection.  In order to restrict the numerical calculations into the
central region $\Omega$ Eq.~(\ref{greenR}) is written in the form
\begin{equation}
\big{(}\omega-\hat{H}_0 -\Sigma_L^r(\omega) - \Sigma_R^r(\omega)\big{)}
G^r(x,x';\omega) = \delta(x-x'),
\end{equation}
where $\Sigma_{L/R}^r(\omega)$ are the self-energies of the leads
$\Omega_{L/R}$ and $\hat{H}_0$ is the Hamiltonian of the isolated
region $\Omega$.

The electron density is obtained from the so-called lesser Green
function $G^<$ by integrating over $\omega$
\begin{equation}\label{ele_integraali}
n_-(x) = \frac{-1}{2\pi} \int_{-\infty}^{\infty} {\rm
Im}(G^<(x,x;\omega)) d\omega.
\end{equation}
When no bias voltage is applied the system is in equilibrium and $G^<$
is calculated as
\begin{equation}\label{ele1}
G^<(x,x';\omega) = 2F_{L/R}(w) G^r(x,x';\omega).
\end{equation}
Above, $F_{L/R}$ are related to the Fermi distributions in
$\Omega_{L/R}$ and in the equilibrium they are equal.  The effects of
the perpendicular directions in our computationally one-dimensional
system are included in $F_{L/R}$ by integrating the Fermi
distributions over $\omega_\bot$
\begin{equation}\label{eq:F}
\begin{aligned}
F_{L/R}(\omega) = \frac{1}{\pi}\int_0^\infty
\frac{1}{1+e^{\frac{\omega+\omega_\bot - \mu_{L/R}}{k_B T}}} d\omega_\bot \\
= \frac{1}{\pi} k_B T \ln \left( 1+
e^{\frac{\mu_{L/R}-\omega}{k_B T}} \right).
\end{aligned}
\end{equation}
When the bias voltage $V_{SD}$ is applied $F_L(\omega)$ and
$F_R(\omega)$ are split by $V_{SD}$ on the energy axis.  In this case
$G^<$ has to be calculated as
\begin{equation}
\begin{aligned}
\label{ele2}
 G^<&(x,x';\omega) = \\ 
& -i F_R(\omega) 
 G^r(x,x_R;\omega) \Gamma_R(x_R,x'_R;\omega)
G^a(x'_R,x';\omega) \, \\ 
&-iF_L(\omega) G^r(x,x_L;\omega)
\Gamma_L(x_L,x'_L;\omega)  G^a(x'_L,x';\omega),
\end{aligned}
\end{equation}
where $x_{L/R}$ are the coordinates of the boundaries
$\partial\Omega_{L/R}$ and $\Gamma_{L/R}$ are defined as
\begin{equation}\label{gammat}
i\Gamma_{L/R} = \Sigma^r_{L/R} - \Sigma^a_{L/R} = 2i {\rm Im}(\Sigma^r_{L/R}).
\end{equation}
Eq.~(\ref{ele2}) is valid also in equilibrium, but only when there
are no bound states. Namely, in this form the electron density in
$\Omega$ is composed of scattering electron states coming from
$\Omega_{L}$ and $\Omega_{R}$.

\begin{figure}[!htb]
\begin{center}
\epsfig{file=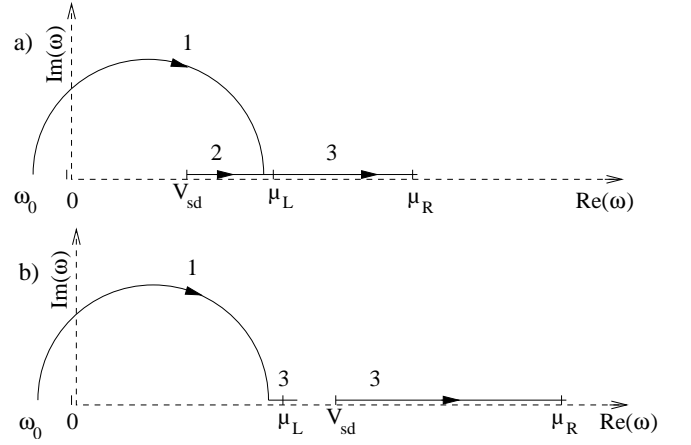,width=0.48\textwidth}
\end{center}
\caption{\label{polku} Integration paths for the electron density
calculations. The paths are divided into three parts $n_{-1}$,
$n_{-2}$ and $n_{-3}$. a) and b) correspond to the cases
$V_{SD}<\mu_L$ and $V_{SD}>\mu_L$, respectively.}
\end{figure}

To perform the energy integral in Eq. (\ref{ele_integraali}) is the
heaviest part of the calculations.  The calculation of $G^<$ at
several energies $\omega$ takes CPU time but cannot be avoided due
to the sharp resonance peaks in our RTD system.  In order to reduce
the number of $\omega$ values needed we move parts of the integral to
the complex plane where the changes in $G^<$ are smooth. The move of
the integration part away from the real axis requires that the
integrand is analytic above the real axis. To fulfill this we first
approximate Eq.~(\ref{eq:F}) as
\begin{equation}\label{appro}
F_{L/R} \approx \frac{\mu_{L/R}-\omega}{\pi}.
\end{equation}
This is exactly true at zero temperature and a good approximation
at energies few $k_B T$ below the Fermi-levels $\mu_{L/R}$.  Now we
can split the integral into three parts $n_{-1}$, $n_{-2}$, and
$n_{-3}$
shown in Fig.~\ref{polku}. The slightly complicated division is due to
the proper inclusion of the perpendicular kinetic energy
component. Here and below we assume that $\mu_{L} < \mu_{R}$. In the
first part $n_{-1}$ the form of $G^<$ in Eq.~(\ref{ele1}) is used and
the integral
\begin{equation}
n_{-1} =  
\int_{\omega_0}^{\mu_L-\Delta k_B T} \frac{1}{\pi}(\mu_L -
\omega)  {\rm Im} \left[G^r(x,x';\omega\right] d\omega
\end{equation}
is moved to the complex plane. Above, $\Delta$ is about 3 ... 5 so
that Eq. (\ref{appro}) is approximately valid. The integral starts at
the energy $\omega_0$, which is below the bottom of the conduction
band of the left lead so that the contribution of possible bound
states is also included. The integral ends at a couple of $k_B T$
below the Fermi-level $\mu_L$.

The integrals $n_{-2}$ and $n_{-3}$ do not vanish in
non-equilibrium. Because Eq. (\ref{ele2}) has to be used the
integrands are not analytic on the complex plane and the integrations
have to be performed along the real axis.  The integral $n_{-2}$
does not vanish when $V_{SD}$ is smaller than the width of the
occupied conduction band in the leads $\Omega_{L/R}$ (the case of
Fig. \ref{polku}a) and is obtained as
\begin{equation}\label{int-2}
\begin{aligned}
n_{-2} =  
&\int_{\omega_0+\mu_R-\mu_L}^{\mu_L- \Delta k_BT}
\frac{1}{\pi}(\mu_R - \mu_L) \times\\
&G^r(x,x_R)\Gamma_R(x_R,x'_R)
G^a(x'_R,x') d\omega.
\end{aligned}
\end{equation}
The integral $n_{-3}$ starts at a couple of $k_BT$ below $\mu_L$
and ends at a couple of $k_BT$ above $\mu_R$. It is obtained as
\begin{equation}\label{int-3}
\begin{aligned}
n_{-3} = 
&
\frac{1}{2\pi}\int_{\mu_L- \Delta k_BT}^{\mu_R+\Delta k_BT} \\
\big[& \,F_R(\omega) \,G^r(x,x_R) \, \Gamma_R(x_R,x'_R)\, G^a(x'_R,x') \\ 
+&\,F_L(\omega) \, G^r(x,x_L) \, \Gamma_L(x_L,x'_L) \, G^a(x'_L,x') 
\,\big]\, d\omega . 
\end{aligned}
\end{equation}
When $V_{SD}$ is larger than the width of the occupied conduction band
in the leads $\Omega_{L/R}$ (the case of Fig. \ref{polku}b) electron
states from $\mu_L+ \Delta k_BT$ to $V_{SD}$ are not occupied.

The electron tunneling probability through $\Omega$ is calculated
using the Green's functions as
\begin{equation}\label{tunneling_probability}
\begin{aligned}
T(\omega) =& \Gamma_L(x_L,x'_L;\omega) \, G^r(x'_L,x_R;\omega ) \\
\times& \Gamma_R(x_R,x'_R;\omega) \,G^a(x'_R,x_L;\omega),
\end{aligned}
\end{equation}
and the current is obtained by integrating over $\omega$, i.e.
\begin{equation}\label{gff:curr}
I = \frac{1}{\pi} \int_{-\infty}^{\infty} T(\omega) \left( \mu_L(\omega) -
\mu_R(\omega) \right) d\omega.
\end{equation}
This whole integral has to be calculated on the real axis, which is
not a problem because the integral is evaluated only once after the
self-consistent solution of the electron density is reached.

\subsection{Finite-element implementation of the GF formalism}

We have implemented the GF formalism using the FEM. For the FEM
implementation we need to write the equations to be solved in the
variational form.  We start from Eq.~(\ref{greenR}) and multiply both
sides by a continuous well-behaving function $v(x)$ and integrate over
$\Omega$. After modifications the equation obtains the form
\begin{equation}
  \label{eq:variational_formulation}
  \begin{aligned}
    \int_\Omega &\Big{\{} - \nabla v(x) \cdot \frac{1}{2} \nabla G^r
    (x,x';\omega) \\ & + v(x) \big{[} \omega-V_{eff}(x)
    \big{]} G^r(x,x' ; \omega) \Big{\}} \,dx\\
    &+ v(x_L)\,\hat{\Sigma}^r_{L}(x_{L},x_{L})\,G^r(x_L,x_L ; \omega)\\
    &+ v(x_R)\,\hat{\Sigma}^r_{R}(x_{R},x_{R})\,G^r(x_R,x_R ; \omega)\\
    &= \, v(x').
  \end{aligned}
\end{equation}
Here the self-energy operators $\hat{\Sigma}^r_{R/L}$ have the
analytic solutions \cite{oma}
\begin{equation}
\hat{\Sigma}^r_{L/R}(x,x') = \frac{1}{4} \frac{\partial^2 g_{L/R}^e
  (x,x' ; \omega)} {\partial x \partial x'},
\end{equation}
where $g_{L/R}^e$ are the retarded Green's functions in the isolated
leads $\Omega_{L/R}$ so that they vanish at the boundaries $\partial
\Omega_{L/R}$. In our model the potential is constant in the leads so
that
\begin{equation}\label{summation}
g_{L/R}^e(x,x') = -\frac{i}{\sqrt{2\omega}} \left(
e^{i\sqrt{2\omega}(x-x')} - e^{i\sqrt{2\omega}(x+x'-2x_{L/R})} \right).
\end{equation}

In the FEM the retarded Green's function is expanded in the basis
$\phi_i$,
\begin{equation}\label{approx}
G^r(x,x') \approx \sum_{i,j=1}^N g_{ij} \phi_j(x) \phi_i(x').
\end{equation}
This is an approximation due to the finite number $N$ of the basis
functions. Eq.~(\ref{approx}) is inserted into
Eq.~(\ref{eq:variational_formulation}). The values of the coefficients
$g_{ij}$ are then calculated by choosing $v(x) = \phi_k(x)$.

We use a basis consisting of the linear functions $\phi^0$ and
$\phi^1$ and of high-order polynomials $\phi^j$ \cite{ele}. That is,
\begin{equation}
  \begin {aligned}
    &\phi^0(\chi) = \frac{1}{2}(1-\chi) \\ &\phi^1(\chi) =
    \frac{1}{2}(1+\chi) \\ &\phi^j(\chi) =
    \sqrt{\frac{1}{2(2j-1)}}(P_j(\chi)-P_{j-2}(\chi)) \quad ,j=2,3,...
 \end{aligned}
\end{equation}
where $P_j(\chi)$'s are the Legendre functions of the order
$j$. $\phi^j(\chi)$'s are given in the reference element with
$\chi=[-1,1]$.  The linear functions span the region of two elements,
whereas the $j>1$ functions are localized within one element only.
The inclusion of the high-order basis functions reduces the number of
basis functions needed in order to obtain accurate results.  This has
a remarkable effect in GF electron structure calculations even for
one-dimensional systems. The calculation of the $G^r(x,x';\omega)$
requires the inversion of a matrix of the size of $N \times N$.  The
derivatives of the functions with $j>1$ are orthogonal to each
other. This makes their use numerically stable and we have implemented
elements up to the fifth order.

\subsection{Wigner function formalism}

We have implemented also the WF formalism for the electron density and
the current calculations.  Reviews of the WF formalism for studies of
open systems can be found in Refs.~\cite{frensley} and
\cite{jacoboni}.  There are similarities between our GF and WF
implementations in the treatment of the one-dimensional equations
resulting from the translational invariance in the directions
perpendicular to the electron current.  They show up, for example, in
the boundary conditions as will be discussed below.

The derivation ~\cite{frensley,jacoboni} of the WF formalism starts by
making a coordinate transformation to the quantum-mechanical density
matrix
\begin{equation}
  \rho(x,x') = \sum_i w_i \langle x | i \rangle \langle i | x' \rangle\,,
\end{equation}
where $x$ and $x'$ are spatial coordinates, $|i\rangle$ is a complete
set of states, and $w_i$ gives the probability of finding a particle
in state $|i\rangle$. The new coordinates, $q$ and $r$, are given by
the relations
\begin{equation}\label{eq:koordinaatit}
\begin{aligned}
&q = \frac{1}{2}(x+x')\\
&r = (x-x')\,.
\end{aligned}
\end{equation}
The WF $f(q,p)$ is now defined as the Fourier transform of the density
matrix, i.e.
\begin{equation}
f(p,q) = \int_{-\infty}^\infty
e^{-ipr}\rho(q+\frac{1}{2}r,\,q-\frac{1}{2}r)dr\,
\end{equation}
and a classical phase-space representation $f(q,p)$ is obtained.

In order to use the WF formalism in transport theory we must study the
time evolution of the WF. Taking the time derivative of the density
matrix and substituting it into the Schr\"odinger equation gives the
quantum-mechanical Liouville--von~Neumann equation
\begin{equation} \label{eq:liouville}
        i \frac{\partial \rho}{\partial t}=
        [ \hat{H},\rho] \equiv \mathcal{L} \rho,
\end{equation}
where $\hat{H}$ is the Hamiltonian of the system and $\mathcal{L}$ is
the Liouville superoperator. Substituting the Hamiltonian $\hat{H}$ of
Eq.~(\ref{hamiltonian}) into Eq.~(\ref{eq:liouville}), we obtain
\begin{eqnarray} \label{eq:liouville2}
        i \frac{\partial \rho}{\partial t} & = & -\,\frac{1}{2}\,
        \left( \frac{\partial ^2}{\partial x^2} - \frac{\partial^2}
        {\partial x^{'2}} \right) \rho \nonumber\\
	& & + \left[V_{eff}(x)-V_{eff}(x')\right] \rho\, ,
\end{eqnarray}
where $V_{eff}$ is the SDFT effective potential given in
Eq.~(\ref{veff}). If a Fourier transform, similar to that for the
density matrix, is performed on Eq.~(\ref{eq:liouville2}) the
transport equation for the WF becomes
\begin{eqnarray} \label{eq:wigner_transport}
        \frac{\partial f(q,p)}{\partial t} & = & -\,p\,
        \frac{\partial f(q,p)}{\partial q} \nonumber\\
	& & -\int_{-\infty}^{\infty} \frac{1}{2 \pi}
        \, V(q,p-p')\,f(q,p')\, dp'\,.
\end{eqnarray}
Above,
\begin{eqnarray}
        V(q,p) & = & 2 \int_{0}^{\infty} \sin(p r)\, \times \nonumber\\
        & & \,\,\,\,\,\,\,\,
	[\,V_{eff}(q+\frac{1}{2}r)
	  -V_{eff}(q-\frac{1}{2}r)\,]\,dr\,.
\end{eqnarray}
is a one-dimensional potential kernel. The effect of $V_{eff}$ is
non-local, incorporating quantum interference effects.  Namely, the
potential kernel $V(q,p-p')$ spreads according to
Eq.~(\ref{eq:wigner_transport}) the WF $f(q,p)$ among different values
of $p$ and adds the interference between alternative paths to the
formulation.

The electron density $n_-(q)$ and current density $J(q)$ are
calculated as
\begin{equation} \label{eq:curr}
        J(q)=\frac{1}{2 \pi}\int_{-\infty}^{\infty} f(q,p) \,p \, dp
\end{equation}
and
\begin{equation} \label{eq:c_dens}
        n_-(q)=\frac{1}{2 \pi}\int_{-\infty}^{\infty} f(q,p) \, dp\,.
\end{equation}

In an open system we need to make a difference between the incoming
and outgoing particles to account for the irreversibility. Our system
is one-dimensional, so that $0\leq q \leq W_\Omega$, and there are two
boundaries at $q=0$ and $q=W_\Omega$. Here, $W_\Omega=x_R-x_L$ is the
width of the central region $\Omega$. Since the characteristics are
first-order equations only one boundary value is needed. Moreover,
since particles which have $p>0$ are moving in the positive direction
on the $q$-axis, we must supply the boundary conditions on the left
hand side boundary $\partial\Omega_L$ from where they are originating
with the momentum distribution in the reservoir. Similarly particles
with $p<0$ are moving in the negative direction on the $q$-axis so
that the boundary conditions must be specified on the right hand side
boundary $\partial\Omega_R$. This scheme invokes the boundary
conditions
\begin{equation}\label{eq:bound_cond}
\begin{aligned}
        f(0,p)|_{p>0}  &=  F_L \left( \frac{1}{2}p^2 \right),\\
        f(W_\Omega,p)|_{p<0}  &=  F_R \left( \frac{1}{2}p^2 \right),
\end{aligned}
\end{equation}
where $F_{L/R}$ are as defined in Eq.~(\ref{eq:F}) and take again the
integration over the perpendicular energy components into account.

We solve the Wigner transport equation for the steady state with
$\partial f/\partial t = 0$ numerically using the discretization
scheme explained in Ref. \cite{frensley}.  The position coordinate $q$
is simply discretized as $N_q$ equally-spaced points with the spacing
$\Delta_q=W_{\Omega}/(N_q-1)$.  Once $\Delta_q$ is fixed the Fourier
completeness relation gives the grid spacing $\Delta_p$ as a function
of $\Delta_q$ as
\begin{equation} \label{eq:discrete_p}
        p_k=\frac{\pi}{\Delta_q}
        \left(\frac{k-1}{N_p}-\frac{1}{2}\right), \,\,
        k=1,2,\, \ldots, N_p\,.
\end{equation}

The derivative of $f(q,p)$ with respect to $q$ in the transport
equation (\ref{eq:wigner_transport}) has been calculated using the
second-order differencing scheme (SDS). I.e.,
\begin{equation} \label{eq:differencing_scheme}
  \begin{aligned}
  &\frac{\partial f(q,p)}{\partial q} = \\
    &\quad \pm  \left[\frac{3  f(q,p) - 
      4 f(q\pm \Delta_q, p) + f(q \pm 2 \Delta_q, p)}
    {2 \Delta_q}  \right],
\end{aligned}
\end{equation}
where the different signs are chosen according to the signs of $p_k$
in such a way that the proper boundary conditions given by
Eq.~(\ref{eq:bound_cond}) are coupled to the transport equation. It
has been shown by Buot and Jensen \cite{jensen,buot} that the SDS
gives considerably better results than the conventional
upwind-downwind differencing scheme (UDS).


\subsection{Differences between the Green's function and Wigner
function formalisms}

The GF and WF formalisms are related. Namely, we can first calculate
the density operator in the GF formalism as
\begin{equation}
\rho(x,x') = -i \int_{-\infty}^\infty G^<(x,x';\omega) \, d\omega.
\end{equation}
Then the coordinate transformation of Eq. (\ref{eq:koordinaatit}) is
performed and we get the WF $f(p,q)$. If $f(p,q)$ is calculated in
this way Eqs.~(\ref{eq:c_dens}) and (\ref{eq:curr}) of the WF
formalism give exactly the same results as Eqs. (\ref{ele_integraali})
and (\ref{gff:curr}) of the GF formalism.

The differences in the results of the two formalisms are caused by the
calculation of $f(p,q)$ in the WF formalism by using the
Liouville--von~Neumann equation. For an infinite system the results
would still be equal but a finite calculation region causes the
differences.  Namely, the WF formalism does not have similar totally
open boundary conditions as the GF formalism. Differences arise in
this case also because there is no energy dependence in the
Liouville--von~Neumann equation whereas $G^<$ depends on $\omega$.

Using the GF formalism typically means that we have to calculate a lot
of information which is of no further use.  The calculation of
$G^R(x,x';\omega)$ using $N$ basis functions means inverting a $N
\times N$ matrix. This requires the solution of $N$ linear equations
each having $N$ unknown variables. The coefficient matrix is sparse,
including only overlapping terms of the basis functions.  In order to
calculate the electron density we have to integrate over $\omega$,
which results in calculating $G^r(x,x';\omega)$ many times.  Luckily,
in the integrals $n_{-2}$ and $n_{-3}$
(Eqs. (\ref{int-2}) and (\ref{int-3})) we need only two linear
equations for $G^r(x_L,x)$ and $G^r(x_R,x)$.

In the WF formalism we need to solve for the Wigner function, which
depends on two variables $p$ and $q$. Because of the use of the
discretization form of Eq. (\ref{eq:differencing_scheme}) the
dependence between these variables is more complicated than that
between $x$ and $x'$ in the GF formalism. This means that we solve a
set of linear equations which includes $N_p \times N_q$ unknown
variables.  The coefficient matrix has a belt type filling where the
width of the belt is large. This makes the linear equation hard to
solve.

\section{Results}

In this section we give results of our electronic structure and
electron current calculations for the magnetic RTD structure described
in Sec. IIA. We concentrate on the feasibility of the GF and WF
formalisms to model this kind of systems.  In the first subsection
this is done by comparing the results of the two formalisms with each
other. In the second subsection we compare our results with those
measured by Slobodskyy {\it et al.} \cite{mit} and thereby analyze the
electronic structure of the actual device. Finally, we make
predictions for the spin-polarized current when structure parameters
of the magnetic RTD are varied.

\subsection{Comparison between the Green's function and 
Wigner function formalisms}

\begin{figure}[!htb]
\begin{center}
\epsfig{file=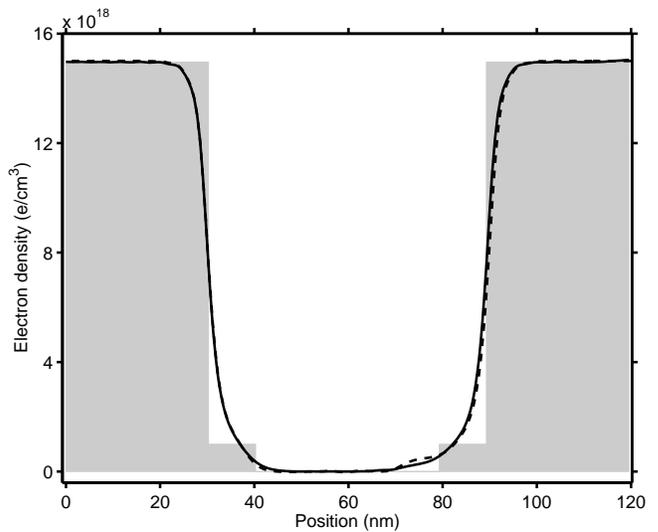,width=0.48\textwidth}
\end{center}
\caption{\label{GWele} Electron density calculated using the GF (solid
  line) and the WF (broken line) formalisms. The values of the
  temperature, magnetic field, and bias voltage used are 4.2~K, 6~T
  and 0.15~V, respectively. The gray areas denote level of positive
  back ground charge.}
\end{figure}

\begin{figure}[!htb]
\begin{center}
\epsfig{file=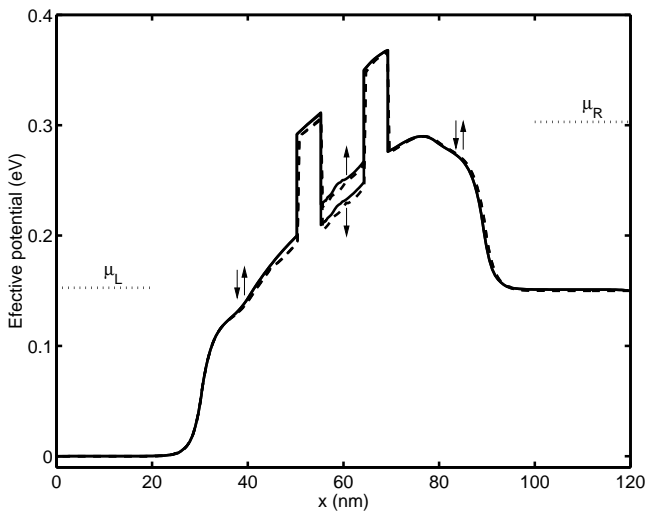,width=0.48\textwidth}
\end{center}
\caption{\label{GWpot} Effective potential calculated using the GF
  (solid line) and the WF (dashed line) formalisms. The values of the
  temperature, magnetic field, and bias voltage used are 4.2~K, 6~T
  and 0.15~V, respectively.  The effective potentials for spin-up and
  spin-down electrons differ remarkably only between the potential
  barriers.}
\end{figure}

The electron density and the effective potential corresponding to the
0.15 V bias voltage and calculated using the GF and the WF formalisms
are shown in Figs.~\ref{GWele} and \ref{GWpot}, respectively. The two
formalisms give very similar results.  There are Friedel oscillations
in the electron density in the leads but their amplitude is so small
that they are not visible on the scale used. The electron density
drops close to zero in the un-doped regions. For this value of
$V_{SD}$ some of the low-energy resonance states are
occupied. However, these states cause only a small density increment
between the potential barriers. The electron density shows in both
formalisms a small asymmetric behavior which becomes clearer with
increasing $V_{SD}$.

The effective potential in Fig. \ref{GWpot} rises strongly at the
interfaces between the doped and un-doped materials. Within the
un-doped region the potential changes then rather linearly due to the
applied bias voltage. The spin-splitting of the potential due to the
external term of Eq. (\ref{deltaE}) does not propagate out of the
quantum well region.  Our GF and WF calculations do not include
inelastic scattering. That would be important for finding
self-consistent solutions if, on the higher potential side of the
barriers, there are ``notch'' states below the band of occupied
electron states \cite{frensley}. In our case ``notch'' states do not
appear because of the strong potential rise due to the un-doped
material layers.

\begin{figure}[!htb]
\begin{center}
\epsfig{file=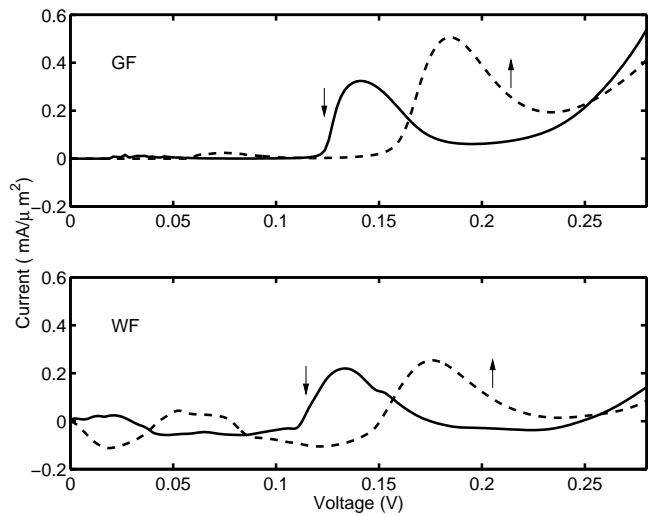,width=0.48\textwidth}
\end{center}
\caption{\label{vertaus} Current through the magnetic RTD structure as
   a function of the bias voltage. The contributions due to spin-down
   (solid line) and spin-up (dashed line) electrons are shown.  The
   results in the upper and lower panels are calculated using the GF
   and the WF formalisms, respectively. The values of the temperature
   and magnetic field used are 4.2~K and 6~T, respectively.}
\end{figure}

The current through our magnetic RTD structure is shown in
Fig.~\ref{vertaus} as function of the bias voltage $V_{SD}$.  The
spin-up and spin-down contributions, split strongly by the magnetic
field of 6~T, show the typical RTD behavior. The most prominent peaks
at around 0.14~V \dots 0.18~V actually correspond to the second lowest
resonance states in energy.  Most of the conduction takes place close
to the right Fermi-level $\mu_R$ because the large un-doped region
diminishes strongly the tunneling probability at lower energies. For
this reason also the current peaks due to the lowest-energy resonances
at around 0.03~V \dots 0.07~V are hardly visible in the GF
results. They are clearer in the WF results, but this is partly due to
numerical difficulties as will explain below.  In the WF formalism the
current may even change its direction and have negative values as seen
in the lower panel of Fig.~\ref{vertaus}.

\begin{figure}[!htb]
\begin{center}
\epsfig{file=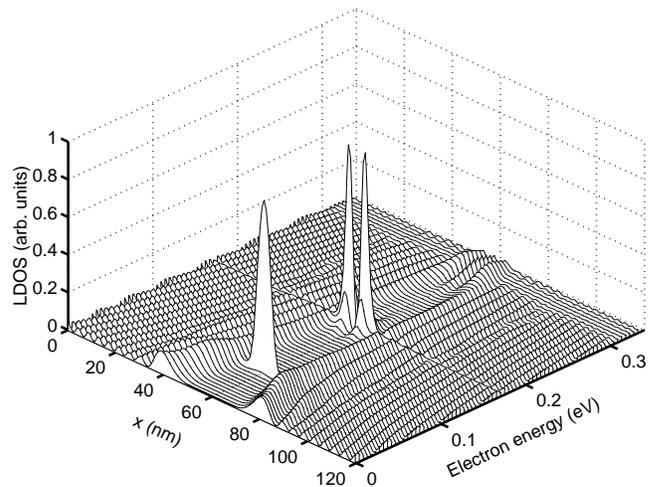,width=0.48\textwidth}
\end{center}
\caption{\label{tila} LDOS of the GF formalism as a function of the
  position $x$ along direction of the current and the energy
  $\omega$. The values of the temperature and magnetic field are 4.2~K
  and 0~T, respectively.}
\end{figure}

The two resonance states per spin in the voltage region of 0 \dots
0.25~V can be clearly seen in the local density of states (LDOS)
calculated in the GF formalism.  Fig.~\ref{tila} shows the LDOS for
the zero-magnetic field and the zero-bias voltage case. In the absence
of the magnetic field, there is no difference between the spin-up and
spin-down electron states.  The first and the second resonance
correspond to the quantum well states with zero and one node plane
parallel to the layer structure, respectively. We can also see that
especially at low energies the LDOS enhances just outside the un-doped
region.

The GF and the WF formalisms give almost equal electron densities and
effective potentials as demonstrated in Figs. \ref{GWele} and
\ref{GWpot}. However, the requirements to reach similar numerical
convergences are very different. In the GF calculations we have used
36 fifth-order elements corresponding to 181 basis functions. A
further increase of the number of elements does not change the
results. In the WF calculations we have used as many as $N_q$ = 320
and $N_p$ = 300 discretization points.  This means that the number of
unknown variables in the linear equations to be solved is of the order
of 10$^5$. If we reduce the number of discretization points from this
magnitude, the effective potential in the RTD region moves down
becoming saggy. The very different numerical characteristics of our WF
and the GF calculations demonstrate the efficiency of the high-order
polynomial basis used in our GF implementation.

In the current-voltage curves of Fig. \ref{vertaus} the differences
between the WF and GF formalisms are more visible than in the electron
density of Fig. \ref{GWele} although the formalisms give the same
qualitative behavior. The resonance peaks of the WF results are
located at slightly lower bias voltages (energies) than those in the
GF results.  This is in accordance with the lower effective potential
in the WF calculations (See Fig. \ref{GWpot}). The GF formalism gives
always by definition a positive current whereas the current calculated
by using the WF formalism may become negative. This is a well-known
artifact of the WF formalism. If we reduce the number of discretation
points the current attains even more negative values. This implies
that the WF results of Fig. \ref{vertaus} are not numerically 
fully convergent. However, with the present computer memory limits we cannot 
increase the number of discretization points much beyond 10$^5$.  At
this stage the WF calculation begins to require also more CPU time
than the GF calculations. This is somewhat surprising because one
would think that the less approximative GF formalism would be
computationally heavier. The use of a more accurate differencing
scheme could improve the WF results without increasing the number of
the discretization points, but this would increase the filling of the
coefficient matrix to be inverted and the CPU time needed.

The numerical problems in the WF calculations are caused by the narrow
resonance peaks. A Fourier transformation has to be evaluated over
electron energies, and the resonance peaks require a remarkable increase of
discretation points. Indeed, Fig.~\ref{vertaus} shows that the current
from the WF calculations for both spins oscillates strongly in the
region of the first resonance peak, which is very narrow. The problem
is not faced in the electron density calculations, because the
contributions of the resonance states to the total electron density
are small. Our real-space GF implementation can handle the narrow
resonance peaks better because we use the adaptive Simpson integral
routine to calculate the density integrals $n_{-2}$, $n_{-3}$, and the
current integral along the real $\omega$ axis and because the integral
$n_{-1}$ is performed in the complex plane where the resonance peaks
are broadened.

\subsection{Comparison to the experiments}

We compare our calculated results with the recent measurements by
Slobodskyy {\it et al.} \cite{mit}. In particular, we have calculated
the current {\em vs.} bias voltage curves using the same magnetic
field and temperature values as they have used.

\begin{figure}[!htb]
\centering
\mbox{\subfigure{\epsfig{file=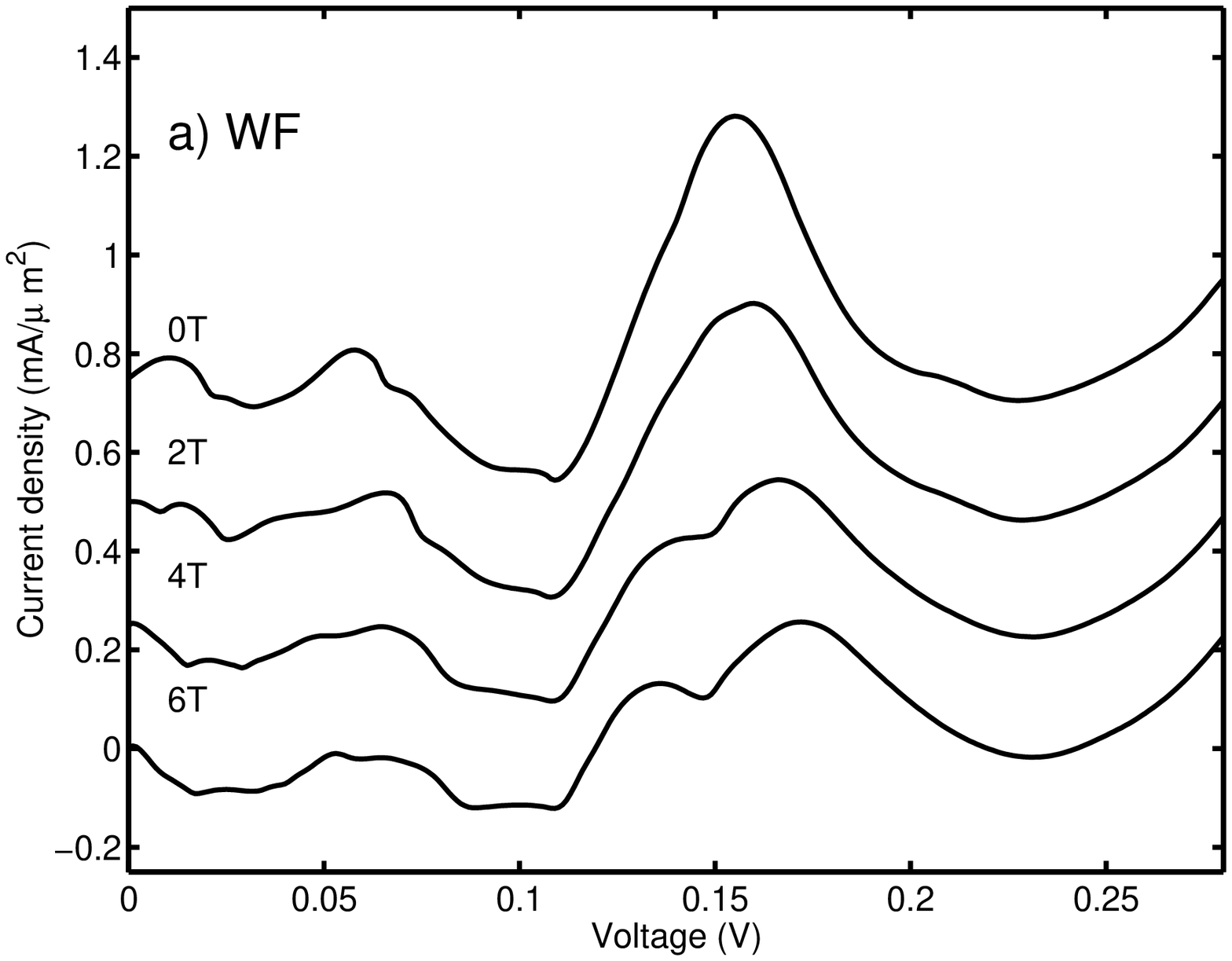,width=0.48\textwidth}}}\\
\mbox{\subfigure{\epsfig{file=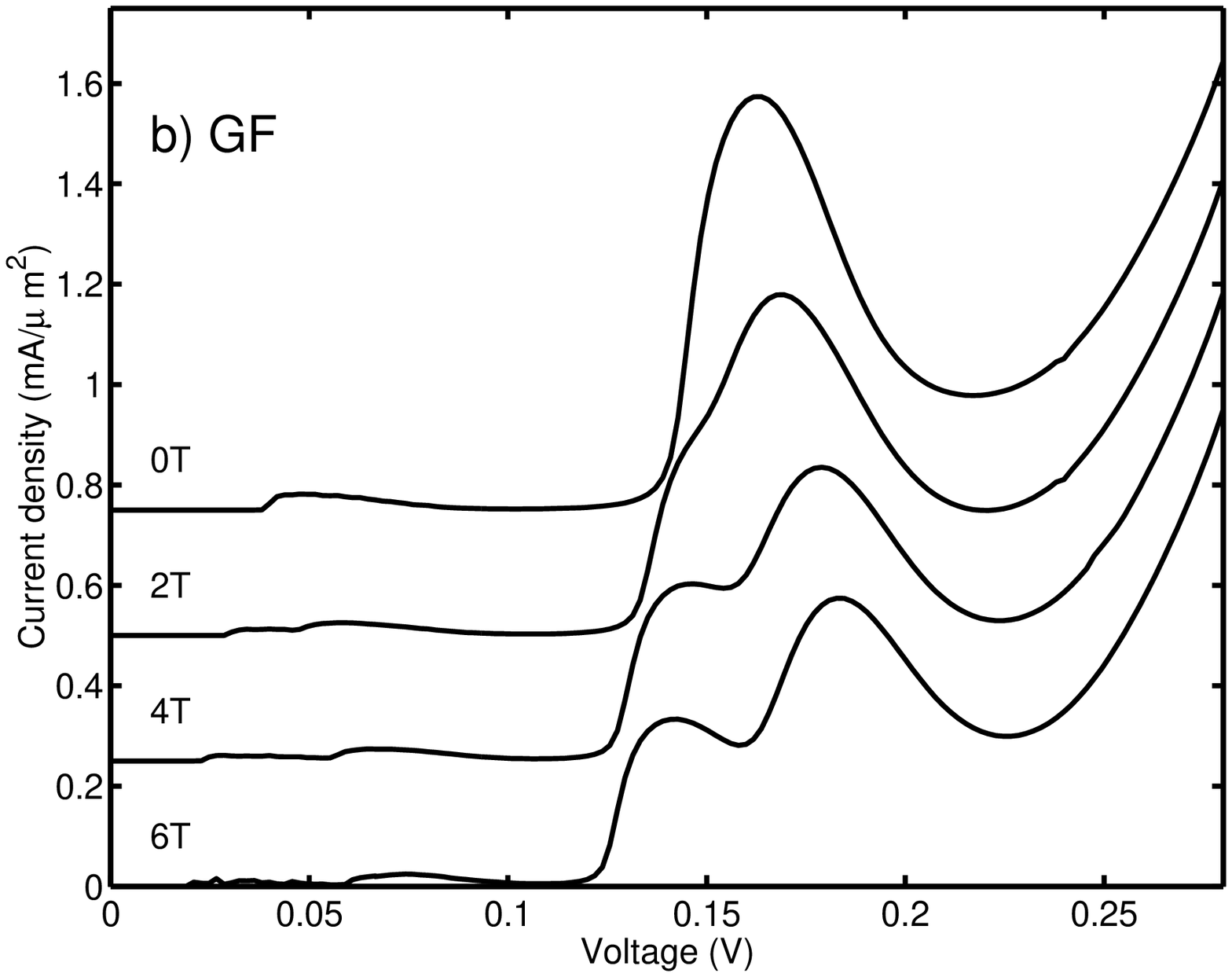,width=0.48\textwidth}}}
\caption{Current through the magnetic RTD structure as a function of
  the bias voltage and different values of the magnetic field.  The
  structure is defined in Fig. \ref{tunneli} and in Table I.  The
  results in the panels a) and b) are calculated using the WF and the
  GF formalisms, respectively. The temperature is 4.2K. For clarity,
  the successive curves are shifted by 0.25~mA/$\mu$m$^2$ with respect
  to each other.}
\label{mag}
\end{figure}

\begin{figure}[!htb]
\centering
\mbox{\subfigure{\epsfig{file=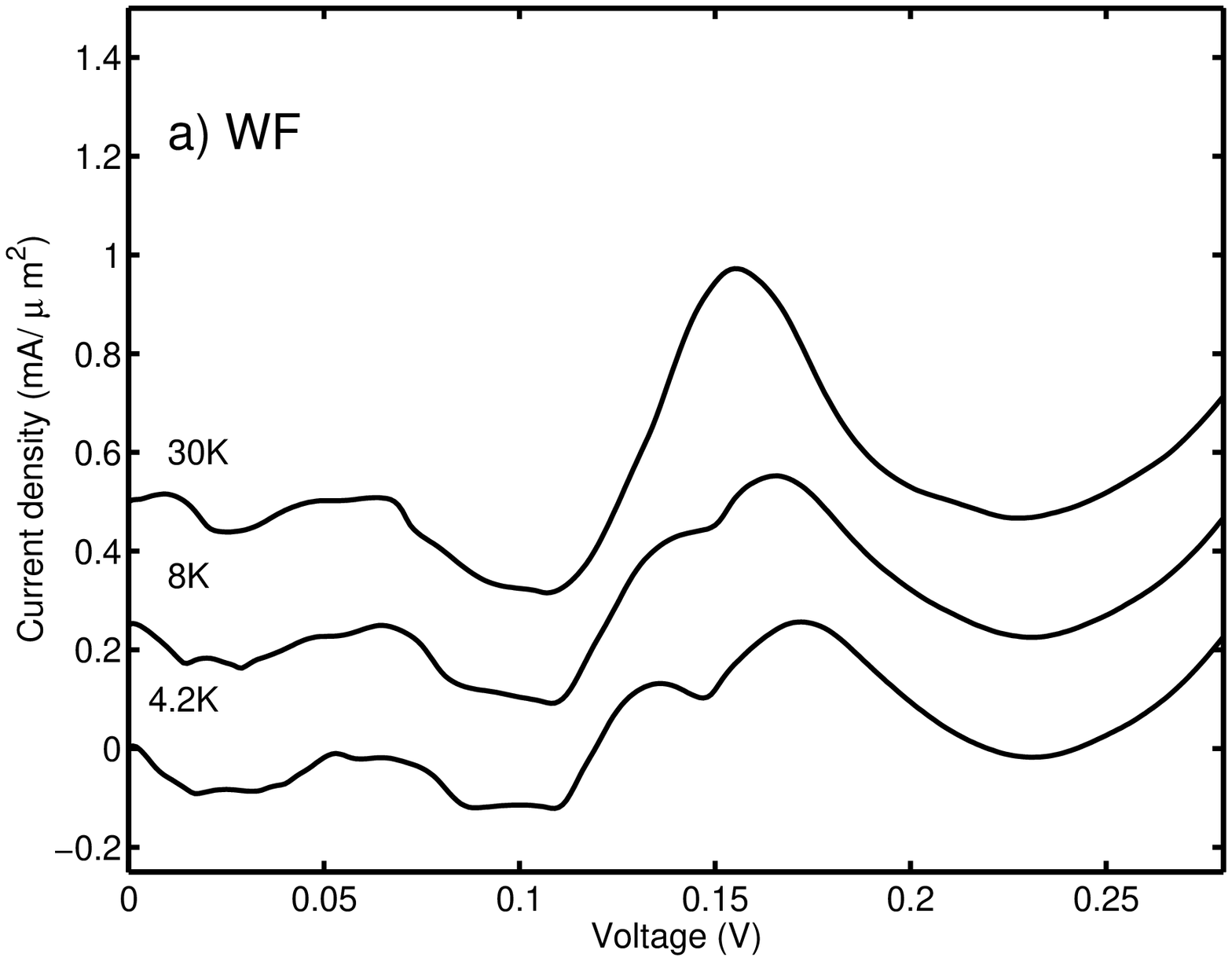,width=0.48\textwidth}}}\\
\mbox{\subfigure{\epsfig{file=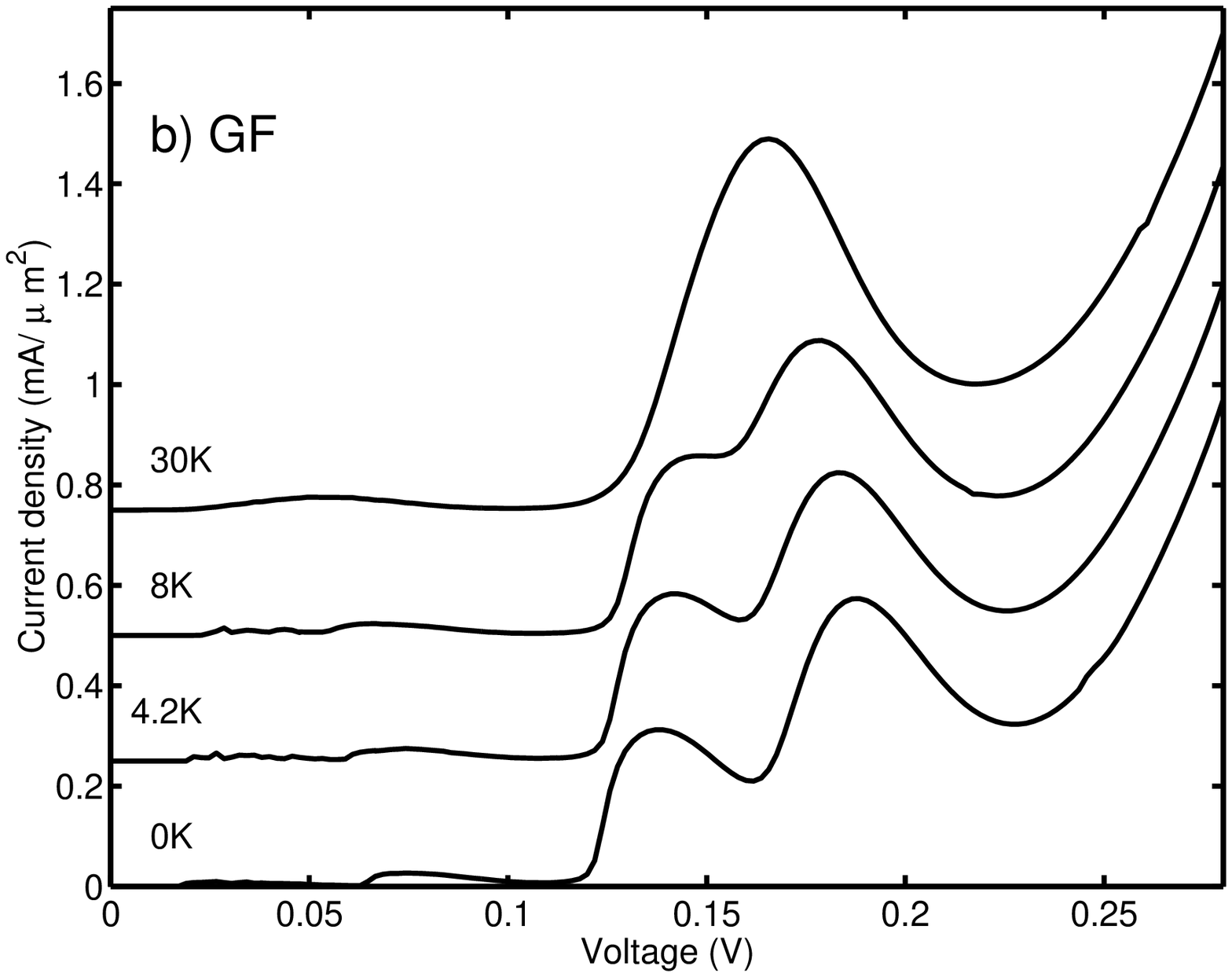,width=0.48\textwidth}}}
\caption{Current through the magnetic RTD structure as a function of
  the bias voltage in different temperatures.  The structure is
  defined in Fig. \ref{tunneli} and in Table I.  The results in the
  panels a) and b) are calculated using the WF and the GF formalisms,
  respectively. The magnetic field is 6T. For clarity, the successive
  curves are shifted by 0.25~mA/$\mu$m$^2$ with respect to each
  other.}
\label{temp}
\end{figure}

The results of the WF and GF calculations for the current as a
function of the bias voltage are shown in Fig.~\ref{mag} corresponding
to different magnetic fields and in Fig.~\ref{temp} to different
temperatures.  As was mentioned above we have chosen the height of the
potential barriers so that the widths of the two prominent spin-up and
spin-down peaks from our GF calculations agree at low temperatures and
high magnetic fields with experiment as well as possible. With this
fitting only, our predictions for the magnetic field and temperature
dependences are rather similar to experiments. The spin-up and
spin-down contributions separate in similar ways with increasing
magnetic field and merge together with an increasing
temperature. These overall features are related to the dependence of
the spin splitting $\Delta E$ on magnetic field and temperature in Eq.
(\ref{deltaE}). One should note that the distances between the spin-up
and spin-down peaks in the current are by a factor of two larger than
$\Delta E$ (See the values in Table II).  The reason is that part of
the voltage is dropped over the first potential barrier. This fact was
taken into account by Slobodskyy {\it et al.}  \cite{mit} by a lever
arm of 2.1 when comparing the measured voltage splitting with the
theoretical $\Delta E$. Our simulations confirm the magnitude of the
lever arm.

A more detailed comparison reveals that in the theoretical
current-voltage curves the resonance peaks are at about 30 mV higher
voltages than in the experimental results. Several effects may
contribute to this difference. First of all, uncertainty in the
thickness of the quantum well affects strongly the resonance
positions.  The value of the series contact resistance in the
measurements, the lack of inelastic scattering and the use of a
constant effective electron mass in our modeling may also shift the
peaks.  However, the energy shifts due to these uncertainties are
expected to be smaller than the distances between the resonances for
each spin direction: the distance between the first and the second
resonance is about 100 mV (see Fig. \ref{tila}). Therefore our
conclusion is that the current peaks seen in the measurements by
Slobodskyy {\it et al.}  correspond to the second resonance state in
energy.

\subsection{Effects of the width of the potential well}

In the previous sections we have concentrated on modeling and
analyzing the magnetic RTD structure by Slobodskyy{\it et al.}  Next
we predict how the geometry of the device affects the spin
polarization of the current. The obvious parameter to be varied is the
width of the quantum well because it determines the positions of the
resonances on the energy axis and affects strongly even the
qualitative features of the current-voltage curves. Thus, in the
following we keep the width and height of the potential barriers the
same as in the previous calculations and vary the width $L_{S5}$ of
the potential well using the values $L_{S5}$ = 0.25$L_{S5}^0$,
0.5$L_{S5}^0$, and 2$L_{S5}^0$, where $L_{S5}^0$ is the original width
of 9 nm (Table I).

\begin{figure}[!htb]
\begin{center}
\epsfig{file=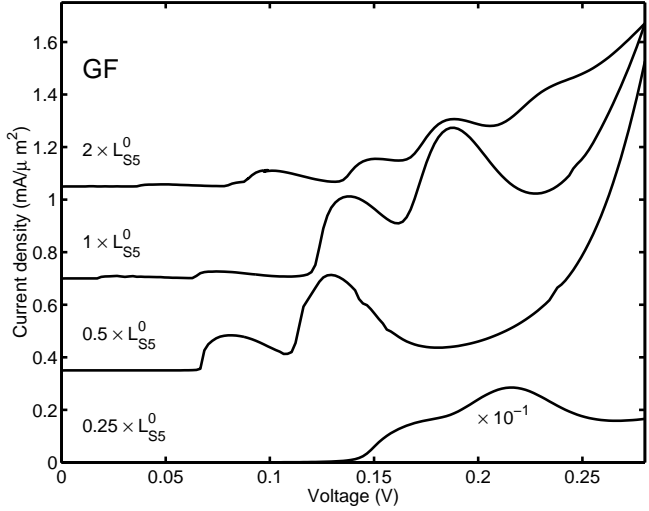,width=0.48\textwidth}
\end{center}
  \caption{Current through a magnetic RTD structure as a function of
  the bias voltage. The structure is defined in Fig. \ref{tunneli} and
  in Table I with the exception that the width of the quantum well
  (region S5) is scaled relative to the original width $L_{S5}^0$.
  The results are calculated using the GF formalism and they
  correspond to the the magnetic field of 6~T and zero temperature.
  For clarity, the successive curves are shifted by
  0.35~mA/$\mu$m$^2$.  The lowest curve for 0.25$\times L_{S5}^0$ is
  scaled by a factor of 1/10.  }
\label{dot_lenght} 
\end{figure}

The current as a function of the bias voltage for the different widths
of the potential well is shown in Fig.~\ref{dot_lenght}.  We see that
the narrowing of the well moves resonance states towards higher
voltages and the first resonance peak for each spin becomes active,
i.e. its contribution to the total current becomes evident. We also
note that the magnitude of the current increases rapidly with
decreasing potential well width. The increase of the potential well
width to 2$L_{S5}$ causes the resonances to come closer to each other
and overlap more strongly. This might be an undesirable feature for
device applications.

\begin{figure}[!htb]
\begin{center}
\epsfig{file=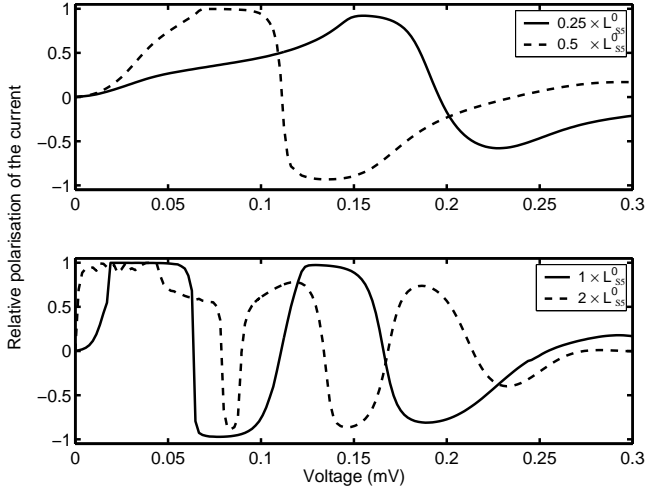,width=0.48\textwidth}
\end{center}
\caption{\label{polarization} Relative polarization of the current
  through a magnetic RTD structure as function of the bias voltage.
  The curves correspond to the total currents shown in
  Fig.~\ref{dot_lenght}.  See also the caption of
  Fig.~\ref{dot_lenght}.}
\end{figure}

Another view of the functioning of the magnetic RTD structure as a
spin switch is given in Fig.~\ref{polarization} showing the relative
polarization of the current, i.e. the difference between the spin-up
and spin-down electron currents divided by the total current. For
structures with very narrow potential wells the polarization is
reduced, although the magnitude of the current is large
(Fig. ~\ref{dot_lenght}). For structures with wide potential wells the
spin splitting for a given magnetic field strength may become
comparable with the distances between the resonances. This causes a
complicated polarization structure as a function of the bias voltage
as can be seen in Fig.~\ref{polarization} for the widest potential
well of 2$L_{S5}$. In conclusion, for a given magnetic field strength
there is an optimum potential well width with good polarization
properties and high current intensity. In the structures studied above
and in the strong field of 6~T it would be approximately 0.3 \dots 0.5
times the original width $L_{S5}^0$.\\


\section{Conclusions}

We have used the semiclassical Wigner function formalism and the
quantum-mechanical Green's function formalism to model spin-dependent
electron transport through a magnetic resonant-tunneling diode
structure. Our calculations are based on the self-consistent solution
of the electronic structures and currents within the
density-functional theory. Both formalisms give very similar results
for the electron density, but the current is more sensitive to the
formalism used. We have traced the differences to be largely due to
the numerical accuracy which is costly to achieve in the
conventional implementation of the Wigner function
formalism. Surprisingly, the Green's function formalism implemented by
using the finite-element method requires less computer resources and
numerically converged results are obtained.
 
Using the two schemes, especially the Green's function formalism, we
have analyzed the recent measurements by Slobodskyy {\it et al.}
\cite{mit} for an actual magnetic resonant-tunneling diode.  The
magnetic-field and temperature dependencies in the measured
current-bias-voltage curves are well reproduced. The two-peak
structure is found to result from the spin-split resonance second
lowest in energy. We show that for a given strength of the magnetic
field causing the spin-splitting there exists an optimum width for the
quantum well so that high spin polarization and current intensity are
achieved.

\begin{acknowledgments}
We acknowledge the generous computer resources from the Center for
Scientific Computing, Espoo, Finland. This research has been supported
by the Academy of Finland through its Centers of Excellence Program
(2000-2005), by the Finnish Cultural Foundation and by the Vilho,
Yrj\"o and Kalle V\"ais\"al\"a Foundation. We are grateful to Per
Hyldegaard, Pekka Kuivalainen and Ville Havu for useful discussions.
\end{acknowledgments}



\begin{thebibliography}{}

\bibitem{mit} A. Slobodskyy, C. Gould, T. Slobodskyy, C. R. Becker,
  G. Schmidt, and L. W. Molenkamp, Phys. Rev. Lett. {\bf 90}, 246601
  (2003).

\bibitem{inj2} H. J. Zhu, M. Ramsteiner, H. Kostial, M. Wassermeier,
H. P. Schonherr, and K. H. Ploog, Phys. Rev. Lett. 87, 016601 (2001).

\bibitem{inj3} V. F. Motsnyi, J. De Boeck, J. Das, W. Van Roy,
  G. Borghs, E. Goovaerts and V. I. Safarov, Appl. Phys. Lett. {\bf
  81}, 2, 265-267 (2002).

\bibitem{inj4} R. Fiederling, M. Keim, G. Reuscher, W. Ossau,
G. Schmidt, A. Waag, and L. W. Molenkamp, Nature (London) 402, 787
(1999).

\bibitem{inj5} Y. Ohno, D. K. Young, B. Beschoten, F. Matsukura,
H. Ohno, and D. D. Awschalom, Nature (London) 402, 790 (1999).

\bibitem{inj1} G. Schmidt, G. Richter, P. Grabs, C. Gould, D. Ferrand,
  and L. W. Molenkamp, Phys. Rev. Lett. 87, 227203 (2001).


\bibitem{frensley} W. R. Frensley, Rev. Mod. Phys. {\bf 62}, 745 (1990).

\bibitem{jacoboni} C. Jacoboni and P. Bordone, Rep. Prog. Phys.
{\bf 67}, 1033 (2004).

\bibitem{datta} S. Datta, {\it Electronic Transport in Mesoscopic
  Systems}, Cambridge University Press (1995).

\bibitem{g1} Roger Lake, Gerhard Klimeck, R. Chris Bowen, and Dejan
 Jovanovic Jour. of Appl. Phys. {\bf 81} 7845-7869, (1997).




\bibitem{g2} K. M. Indlekofer, J. Lange, A. F\"orster, and
H. L\"uth,Phys. Rev. B {\bf 53}, 7392-7402 (1996).


\bibitem{g3} Gianluca Stefanucci and Carl-Olof Almbladh
Phys. Rev. B {\bf 69}, 195318 (2004).


\bibitem{w1}Unlu M. B., Rosena B., Cuia H.-L. and Zhaob P.
Phys. Lett. A {\bf 327}, 230-240 (2004).

\bibitem{oma} P. Havu, V. Havu, M. J. Puska, and R. M. Nieminen,
  Phys. Rev. B {\bf 69}, 115325 (2004).

\bibitem{oma2} P. Havu, M. J. Puska, R. M. Nieminen, and V. Havu, 
unpublished.

\bibitem{me} S. Lee, F. Michl, U R\"ossler, M. Dobrowolska and
  J. K. Furdyna, Phys. Rev. B {\bf 57}, 9695 (1998).


\bibitem{me2} Landolt-B\"ornstein, ed. by O. Madeling, New Series Group
  III, Vol 22, Pt. a (Springer, Berlin, 1987).











\bibitem{aukko-offset} M. Kim, C. S. Kim, S. Lee, J. K. Furdyna,
  M. Dodrowolska, J. of Crystal Growth {\bf 214/215}, 325-329 (2000).

\bibitem{aukko-offset2}C. Chauvet, E. Tournie and J.-P. Faurie,
  Phys. Rev. B, {\bf 61}, 5332 (2000).

\bibitem{xc} D.~M.~Ceperley and B.~J.~Alder, Phys. Rev. Lett. {\bf
45}, 566 (1980);
J. P. Perdew, A. Zunger, Phys. Rev. B {\bf 23}, 5048-5079 (1981).

\bibitem{poisson} J. Arponen, P. Hautoj\"arvi, R. Nieminen, and
  E. Pajanne. Phys. F: Metal Physics, {\bf 3} (12):2092, (1973).


\bibitem{ele} C. Schwab, {\it p- and hp- Finite Element Methods
  -Theory and Applications in Solid and Fluid Mechanics}, Clarendon
  Press, (1998).

\bibitem{jensen} K. L. Jensen and F. A. Buot, J. Appl. Phys. {\bf 67},
  2153 (1990).
  
\bibitem{buot}  F. A. Buot and K. L. Jensen, Phys. Rev. B {\bf 42},
  9429 (1990).


\end{thebibliography}
\end{document}